\documentclass[aps,preprint,groupedaddress,longbibliography]{revtex4-2}

\usepackage{placeins}
\usepackage{amsmath}
\usepackage{graphics}
\usepackage{cleveref}
\usepackage{siunitx}
\usepackage{xcolor}
\newcommand{\dist}{\hspace{5pt}}

\begin{document}
\title{Time-resolved velocity and pressure field quantification 
in a flow-focusing device for ultrafast microbubble production}

\author{Sarah Cleve$^1$}
\author{Christian Diddens$^1$}
\author{Tim Segers$^{2,1}$}
\author{Guillaume Lajoinie$^1$}
\author{Michel Versluis$^1$}
\affiliation{$^1$Physics of Fluids group, MESA+ Institute for Nanotechnology and 
Technical Medical (TechMed) Center, 
University of Twente, P.O. Box 217, 7500 AE, Enschede, THE NETHERLANDS \\ 
$^2$BIOS Lab-on-a-Chip group, Max-Planck Center Twente for Complex 
Fluid Dynamics, MESA+ Institute for Nanotechnology, University of Twente, 
P.O. Box 217, 7500 AE, Enschede, THE NETHERLANDS}

\date{\today}

\begin{abstract}
Flow-focusing devices have gained great interest in the past decade, due to 
their capability to produce monodisperse microbubbles
for diagnostic and therapeutic medical ultrasound applications. 
However, up-scaling production to industrial scale requires
a paradigm shift from single chip operation to highly parallelized systems.
Parallelization gives rise to fluidic interactions between 
nozzles that, in turn, may lead to a decreased monodispersity. 
Here, we study the velocity and pressure field fluctuations in a single 
flow-focusing nozzle during bubble production.
We experimentally quantify the
velocity field inside the nozzle at $\SI{100}{\nano\second}$ time resolution,
and a numerical model provides insight into both the oscillatory 
velocity and pressure fields. 
Our results demonstrate that, at the length scale of the flow focusing channel, the 
velocity oscillations propagate at fluid dynamical time scale 
(order of~$\si{\micro\second}$) whereas the dominant pressure 
oscillations are linked to the bubble pinch-off and propagate 
at a much faster time scale (order of~$\si{\nano\second}$).
\end{abstract}

\maketitle

\section{Introduction}

Coated microbubbles are currently investigated 
for a broad variety of medical applications,  
both in the fields of diagnostic imaging and therapy \citep{unger2004therapeutic},
which include sonoporation \citep{lentacker2014understanding}, 
blood-brain barrier opening \citep{hynynen2001noninvasive} 
and sonothrombolysis \citep{porter2001ultrasound}.
The diagnostic application relies on the 
scattered nonlinear echo
and therapy requires a mechanical response of the
microbubbles to the driving ultrasound.
Both the echo and volumetric oscillation amplitude 
depend on the driving frequency with respect to the 
resonance frequency of the bubble \citep{versluis2020ultrasound} 
which is inversely proportional to the bubble size.
Clinically available ultrasound contrast agents contain
microbubbles with a broad size distribution, typically 
ranging from $1$ to $\SI{10}{\micro\meter}$ in diameter
\citep{frinking2020three}.
A narrowband ultrasound wave typically used for therapy and diagnostics
would therefore effectively drive only a small fraction of 
the contrast agent population contained in the injected dose.
This suboptimal match between ultrasound spectral content and 
bubble size distribution motivates the development and use of monodisperse microbubbles. 
The increased sensitivity of monodisperse bubbles with respect
to polydisperse ones has first been demonstrated for 
molecular imaging using low bubble concentrations \citep{talu2007tailoring}
and later for contrast imaging \emph{in vitro}
\citep{segers2018monodisperse} and \emph{in vivo} 
\citep{helbert2020monodisperse}. In particular,
\citet{segers2018monodisperse} report sensitivity increase
by two and three orders of magnitude for the fundamental and 
second harmonic, respectively.

In order to obtain a monodisperse contrast agent, one can 
either extract bubbles of the desired size from 
a polydisperse suspension, e.g.\ by acoustic 
sorting \citep{segers2014acoustic} and differential centrifugation
\citep{feshitan2009microbubble},
or one can create 
monodisperse bubbles directly, e.g.\ in a microfluidic device. 
In the past two decades, different microfluidic techniques
have been investigated
for the production of monodisperse bubbles
including cross-flow, co-flow and 
flow-focusing as summarized in \citep{rodriguez2015generatio}.
The use of flow-focusing is particularly promising 
as it allows for production rates
exceeding 1\,million bubbles per second \citep{segers2016stability,van2021feedback}.
Flow-focusing devices can be operated in three different 
regimes \citep{sullivan2008role}: 
(1)~In the squeezing regime, also called geometrically controlled break up,
bubbles completely block the outlet channel or orifice before pinch-off. 
(2)~In the dripping regime a gas jet is formed but its tip retracts
periodically from the outlet channel. 
(3)~In the jetting regime, the gas jet stably extends into the flow-focusing channel,
and bubbles pinch off from its tip. 
Among these regimes, the jetting regime is characterized by the highest 
production rates.
However, even at a production rate of 1\,million bubbles per second a single nozzle can 
only produce a few clinical doses per hour.
In order to reach industrially relevant
quantities, upscaling is necessary. 

\bigskip 
Upscaling by replicating the driving and control systems necessary to drive 
a single chip is prohibitive in terms of cost, operation difficulty and space. 
An evident approach to upscaling is therefore the parallelization of multiple
nozzles.
Such a strategy has been investigated for parallel droplet 
production \citep{conchouso2014three,jeong2015kilo}, 
which has then been translated to upscaled bubble production with some success, using 
two to several hundred nozzles 
\citep{hashimoto2008formation,jiang2010mass,chen2011parallel,kendall2012scaled,jeong2017liter,jeong2019large}.
Despite the impressive achievements on the production rate 
\citep{jeong2017liter,jeong2019large} 
the optimal parameter range for stable bubble production
can hitherto only be extracted from empirical methods. 
Furthermore, the different parallelization
strategies explicitly \citep{jiang2010mass,kendall2012scaled}
or implicitly show a loss of monodispersity as compared to a single nozzle system.
In comparison, monodispersity seems to be less affected
in parallel droplet production \citep{hashimoto2007synthesis}. 
Possible reasons are the compressibility of the
gas phase but also the fact that the liquid flow is typically mass flow controlled
while the gas flow is mostly pressure controlled.
In studies where the supply lines were connected \citep{li2008simultaneous},
only weak interactions were reported for droplet production.

The increased polydispersity during bubble production is suspected to arise from the 
communication between nozzles \citep{hashimoto2007synthesis} trough the
common liquid and gas supply lines, as well as through as the common outlet channel.
Two possible sources of communication between channels have been 
discussed in literature \citep{jeong2017liter,hashimoto2007synthesis}.
The first one consists in purely geometrical considerations
such as varying hydraulic resistances of individual channels due to, e.g.\, design or
fabrication inaccuracies.
These geometrical differences can be alleviated by a thorough design
of a completely symmetric
system \citep{chen2011parallel,kendall2012scaled}, or by 
adapting the size of each channel to ensure equal flow resistances for each 
nozzle \citep{jeong2017liter}.
The second type of communication, the so-called 'cross-talk', 
refers to oscillatory flow phenomena in one nozzle propagating to
the neighboring nozzles 
and is more difficult to eliminate or control, \emph{a~priori}. 
Evidence for cross-talk during bubble production in the squeezing regime has been reported by 
\citet{hashimoto2007synthesis,hashimoto2008formation}.
Their system has a common outlet channel and separate, decoupled inlets
for two to four nozzles. There, bubbles in neighboring nozzles are produced in anti-phase. 
The authors argue that a change in flow resistance during bubble production 
directly influences the flow rate provided to the other channel(s), which then leads 
to an alternating bubbling regime.

\bigskip

Crosstalk between
parallelized channels due to flow and pressure fluctuations at the time scale 
of microseconds
seems a likely candidate for influencing the monodispersity of 
the produced bubbles.
However, studies 
have so far only been conducted for the squeezing regime. 
The role of time-resolved cross-talk between parallelized channels
in the jetting regime, has not yet been studied,
to the best of our knowledge. 
However, some existing studies have considered the effect that
externally imposed oscillations have on the production of bubbles.
The influence of an acoustic pressure wave inside the gas 
supply line has been studied by 
\citet{shirota2008formation,shirota2008formationB} for the pinch-off 
of submillimetric bubbles into a quiescent liquid. 
\citet{shirota2008formation} argue that a temporarily increased 
gas inflow caused by the short ultrasound pulse will
promote the bubble pinch-off, while \citet{shirota2008formationB} 
make the negative pressure pulse responsible for this.
Even though the time and length scales in our flow-focusing device 
differ by several orders of magnitude compared to the system used by 
\citet{shirota2008formation,shirota2008formationB}, it is likely that pressure 
pulses do have an influence on bubble-production. 
To  the best of our knowledge, no studies have investigated 
the effect of forced oscillations of the liquid flow on the timescale 
of the production frequency of bubbles.
Such studies have however been 
conducted for droplets, numerically by \citet{mu2018numerical} and experimentally 
by \citet{yang2019manipulation}. The authors find that the 
droplet production frequency will lock into the actuation 
frequency as long as the two are not too different, in that case
the production becomes polydisperse but periodic (binodal, Hopf bifurcation)
or even completely 
unpredictable, probably chaotic. Furthermore, actuating both inner and outer 
liquid during droplet-production has an effect on droplet size 
\citep{mu2018numerical,yang2019manipulation}. 
If velocity fluctuations are strong enough to play 
a role in the coupling between several flow-focusing 
nozzles for bubble production, one could expect a favorable 
locking to one and the same frequency, although 
the exact phase behavior may be both,
either beneficial of unfavorable.
Another type of study \citep{mutlu2018oscillatory,vishwanathan2021inertial}
has shown that oscillations would let particles migrate towards equilibrium positions
inside the channel, however typically requiring time scales much larger
than available in the present system.
\bigskip

Before analyzing typically complex
parallelized systems, we propose to understand the nature and magnitude of the 
oscillatory components in a simpler device with a single flow-focusing nozzle.  
The unsteady nature of the bubbles is easily observed from time-dependent
gas-liquid interface, but fluctuations also propagates through the gas and
liquid phases.
Understanding the unsteady components may aid the development of more
complex systems with parallel flow-focusing nozzles.
The present
study characterizes the pressure and velocity field in a single flow-focusing
nozzle. Ultra-high-speed imaging measurements 
combined with particle tracking velocimetry give access 
to the velocity field, the experimental methods are introduced in
section~\ref{sec_experiments}. 
Numerical simulations, section~\ref{sec_numerics}, confirm these findings
and further allow the detailed study of the dynamic pressure field.
For both velocity and pressure, the field can be separated into
(1) a time-averaged field, section~\ref{sec_average}, 
which can be compared to simplified theoretical models,
and (2) oscillatory components, section~\ref{sec_oscillations},
which reflect the unsteady nature of the bubble-production.
We show that the liquid velocity oscillations are induced 
by an oscillating gas flow rate and the periodic bubble pinch-off, and that 
they propagate at fluid dynamics time scale of microseconds. 
Pressure oscillations on the other hand, are dominated
by a strong pressure pulse linked to bubble pinch-off and propagate
through the flow-focusing region at the timescale of nanoseconds.
A discussion, section~\ref{sec_discussion}, and conclusion,
section~\ref{sec_conclusion}, conclude this manuscript.

\section{Experiments\label{sec_experiments}}

\subsection{Experimental setup\label{sec_setup}}

\begin{figure*}
 \includegraphics{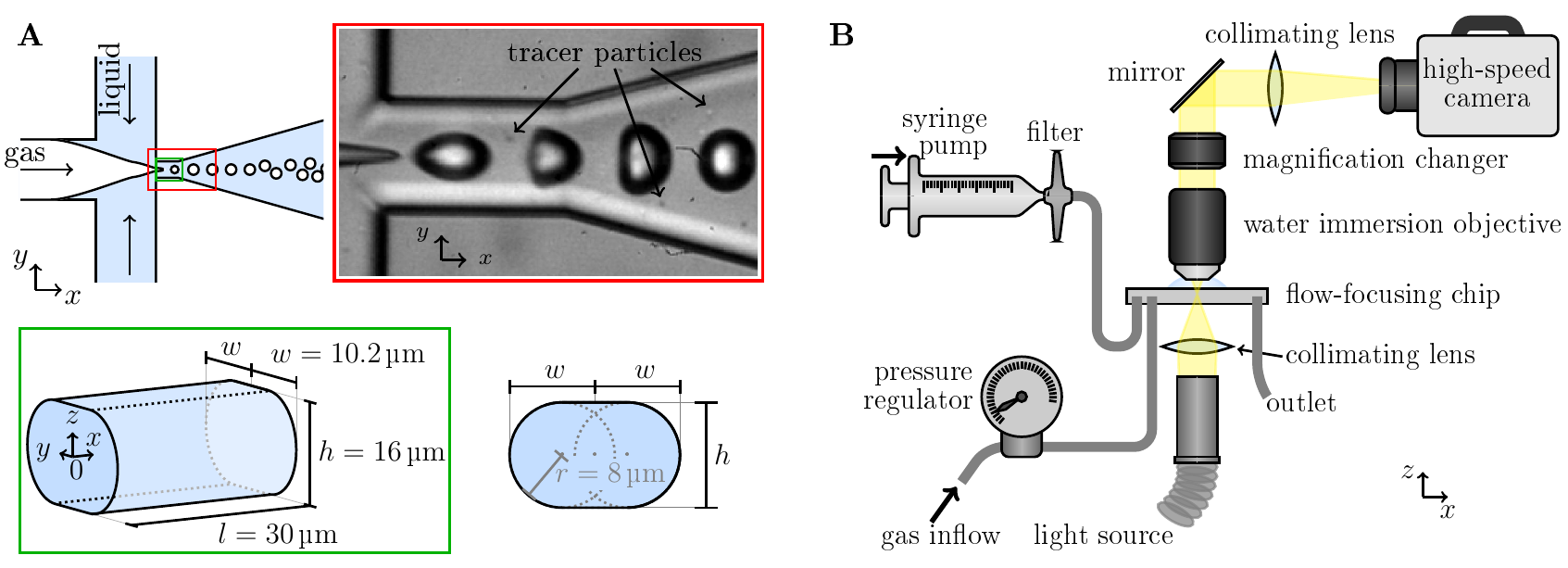}%
  \caption{\textbf{A} -- Schematic of the flow-focusing region with typical snapshot from
   a high speed recording and explanation of the geometry of the flow-focusing channel
   and its cross-section.
   \textbf{B} -- Schematic of the experimental setup. 
\label{fig_setup}}
\end{figure*}

Experiments were conducted on a flow-focusing device that was 
fabricated by isotropic etching of the channel features in 
two glass wafers (i.e. 
with hydrofluoric acid etching at the same rate in all directions
and thus creating corners with a radius equal to the etching depth
\citep{koehler2008Etching}). These were then aligned and bonded 
together as described in \citet{segers2018high}. 
The channel section where the bubbles are being formed, see \cref{fig_setup}\,A,
has a width $b=2w=\SI{20.4}{\micro \meter}$ and height $h=2r=\SI{16.0}{\micro 
\meter}$. The length of this section is $l=\SI{30.0}{\micro\meter}$.

The liquid used in the present experiments was water 
with a weight fraction of 2\,\% Tween\,80 (Sigma-Aldrich).
The surfactant was added to decrease 
the probability of 
bubble coalescence downstream in the expanding outlet \citep{segers2017universal}.
This allowed a larger parameter range for stable
bubble production.
In the nozzle region, Tween\,80 has a similar stabilizing effect on the
bubble production as phospholipids that are typically used
to stabilize medical ultrasound contrast microbubbles.
Tracer particles (red fluorescent microspheres,
$\SI{0.52}{\micro \meter}$ in diameter, Thermo Scientific) were added
with a weight fraction of approximately 
1\,\% to visualize the flow field. 
The liquid was supplied to the chip by a syringe pump (Harvard) 
operated at a constant flow rate. A filter 
(Acrodisc PSF Versapor $\SI{10}{\micro\meter}$, Pall) 
mounted between the pump and the chip was used 
to prevent clogging by dust particles.
The employed gas was air, provided by a laboratory internal pressure supply. A 
pressure regulator (IMI Norgren) was used to control the input gas pressure.
The chip was connected to the inlets via a chip holder (Fluid Connect PRO, Micronit), 
and the nozzle region was visualized using a bright-field modular microscope 
(Olympus, BXF) 
equipped with a $60\times$ water-immersion 
objective (LUMPlanFl, $0.90$ NA, Olympus), an additional $2\times$ magnification unit
(U-ECA, Olympus) and a mirror (U-DP, Olympus). The total imaging magnification was  
$120\times$. A small water drop was placed on top of the chip for 
optical coupling between the chip and the water-immersion objective.
High-speed recordings were acquired with an ultra-high-speed camera (Shimadzu HPV-X2) 
operated at a frame rate of $10$ million frames per second and a resolution of 
$400\times \SI{250}{pixels}$.
The imaging resolution was $\SI{0.25}{\micro \meter /pixel}$, which was 
calibrated with a USAF 1951 calibration grid. 
A Xenon strobe light 
(Vision Light Tech), 
synchronized with the camera and coupled through an optical 
fiber and condensing lens, provided sufficient light intensity for the optical 
recording. 
Each recording allowed to save 256 images.
However, owing to the short duration of
the light pulse of $\sim\SI{15}{\micro\second}$, only about half of 
these recorded images were bright enough for further processing.

\subsection{Experimental procedure}

The chip can be operated at different pressures and liquid flow rates.
In the present paper, we focus on flow rates ranging from
$\SI{150}{\micro \liter \,\minute^{-1}}$ 
to $\SI{250}{\micro \liter \,\minute^{-1}}$ 
and pressures ranging from $\SI{1.8}{\bar}$ to $\SI{3.2}{\bar}$. 
For each measurement series, the parameters were fixed and the chip was 
operated for several minutes in order to ensure a stable bubble production. 
In order to have sufficient sample data to evaluate the mean
velocity field via particle tracking, each measurement series consisted 
of at least 100 recordings.
Subsequently, 400 recordings for each parameter set were saved to extract the
time-resolved velocity field.
Due to the data transfer time of the camera, 
the total duration of a single measurement series ranged from $15$ to 
$\SI{80}{\minute}$.

\subsection{Bubble detection}

The particles and noise were filtered out by means of
singular value decomposition (SVD) by only keeping the large 
coherent structures of the image series. The bubble shapes
were then extracted by applying a threshold.
All extracted information (bubble size, shape and position)
were referenced to the previous pinch-off event so that 
all temporal information is contained within the time span
$[t=0,t=1/f_b = T_b]$, where $f_b$ and $T_b$ 
are the bubble production rate and period, respectively.
The bubble production rates obtained for the present data range from 
$0.65\cdot 10^6$ to $1.3\cdot 10^6$ bubbles per second, with radii ranging from 
$\SI{3.5}{\micro \meter}$ to $\SI{6.0}{\micro \meter}$. 
The bubble volume was obtained through volume 
integration of the bubble contour assuming axial symmetry along the $x$-axis.

\subsection{Particle tracking and flow velocimetry}

\begin{figure*}
\centering
\includegraphics{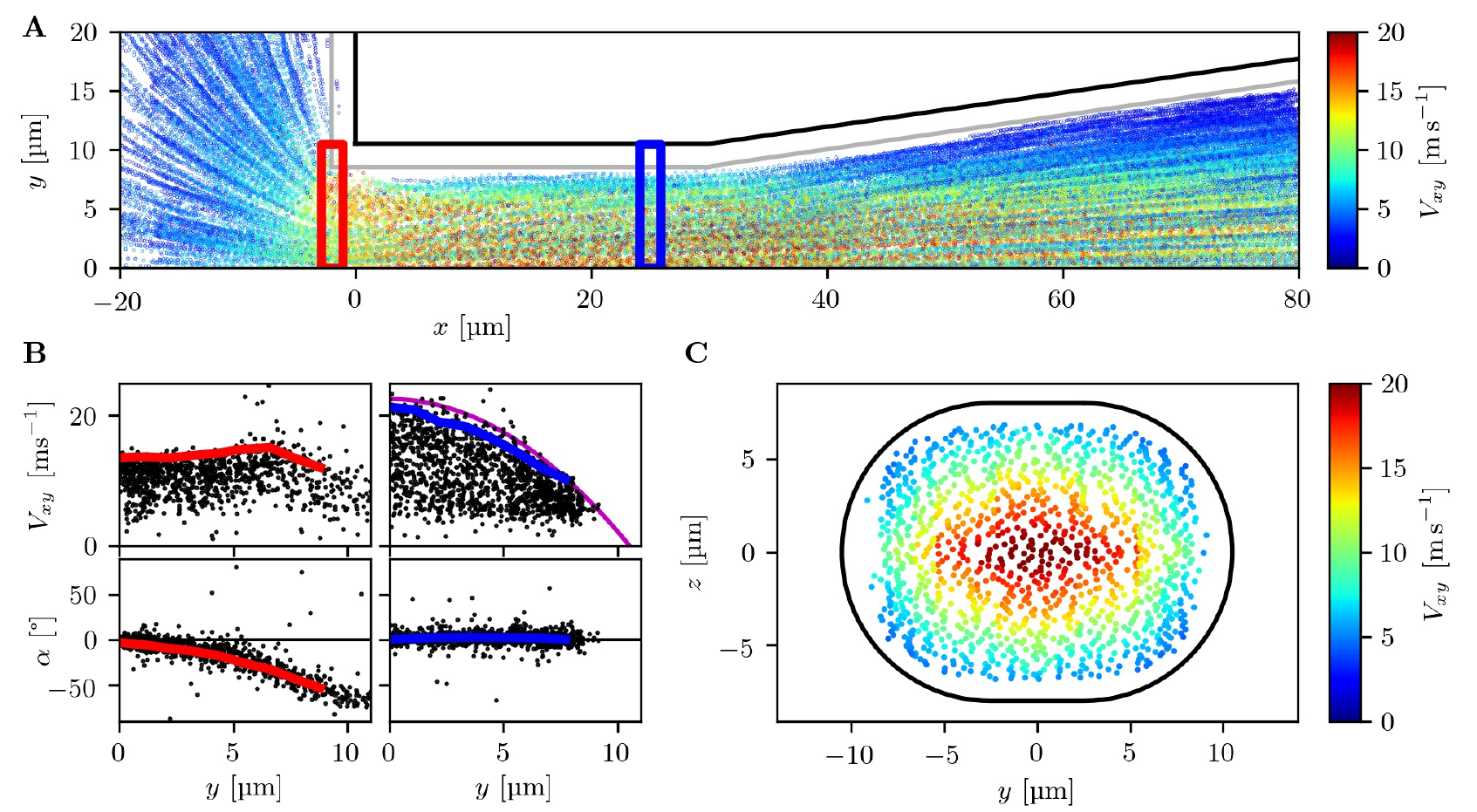}
\caption{
Visualization of the post processing.
\textbf{A} -- More than 130,000 data points for the tracer particles 
presented along the channel profile. 
The total particle velocity $V_{xy}$ is indicated by the color code.
The gray line indicates the border of the shadow region of the channel
wall, where no particles can be detected.
\textbf{B} -- Total velocity $V_{xy}$ and flow direction
$\alpha$ for the two sections marked by the 
rectangles in \textbf{A}. The colored bold lines corresponds to the velocities and
angles obtained through post-processing for the 2D imaging plane. 
The thin, magenta line indicates a Poiseuille flow for comparison.
\textbf{C} -- Reconstruction of the velocity in the cross-section 
$x\approx\SI{25}{\micro\meter}$, based on the assumption that the velocity is maximum
in the channel center and that detected tracer particles are equally spaced
in the $z$-direction for a given position $y$. 
\label{fig_postprocessing}}
\end{figure*}

SVD filtering keeping only small coherent structures and subsequent
noise filtering and thresholding were applied to extract
the positions of the particles.
The positions were then developed into trajectories with the help of the Matlab 
script \emph{track} \citep{crocker1996methods}.
The velocities were subsequently calculated from the temporal 
difference of the particle positions between successive snapshots.
As for the bubbles, all information was related to the bubbling period~$T_b$. 
An example of all extracted sets of information for a flow without bubbles,
totaling over 130,000 data points,
is shown in \cref{fig_postprocessing}\,A.
The large variation in tracer velocities at a single 
location $(x,y)$ reflects the 3D structure of the flow that has 
a low velocity near the top and bottom walls ($|z|\approx r$) and a maximum 
velocity in the channel center ($z\approx 0$).
The theoretical depth of field (DOF) of the 
$60\times$ objective is limited to $\pm\SI{1.6}{\micro\meter}$.
Nonetheless, 
out-of-focus tracer particles located at much larger distances 
from the focal plane are also captured as blurred entities.
The recordings thus gather information on the 3D velocity field. 
While the experimental data does not allow a direct evaluation 
of the $z$-position, we can reconstruct the data as a function of $z$ 
by assuming that (1) velocity is maximum in the center of 
the channel and (2) that the particles are equally distributed in the flow, 
see \cref{fig_postprocessing}\,C.
The main interest for the present study lies in extracting
the velocity field in the central imaging plane ($z=0$), where bubbles are passing.
An example is shown in \cref{fig_postprocessing}\,B for two-channel
cross-sections.
The maximum velocity, assumed to lie in the central plane of the flow
can thus be recovered by taking the envelop of the velocity-data.
We extract the flow-direction,
expressed by the angle $\alpha$, using average values at
every channel position since no
significant dependency of $\alpha$ on the $z$-direction 
was observed, nor expected.
The resulting envelopes shown in \cref{fig_postprocessing}\,B
match the theoretical expectations, i.e., that the flow 
will develop from a flat velocity profile in the entrance region
into a parabolic profile further downstream.
Within the experimental error margin, $\alpha=\arctan(V_y/V_x)\approx0$ everywhere except 
at the entrance of the channel and will therefore have little importance.
In the following, we will thus focus the discussion on the velocity magnitude
$V_{xy} = \sqrt{V_x^2 + V_y^2}$.
Here, $V_x$ and $V_y$ are 
the velocity components in the $x$- and $y$-directions, respectively. 
The example presented in \cref{fig_postprocessing} 
depicts a flow without any bubbles, and it is therefore stationary.
Furthermore, typical flow velocities of \SI{20}{\meter\per\second} 
assure Reynolds numbers of the order of 400 and consequently a laminar flow regime.
The presence of bubbles will add a periodic component in time. 
Post-processing can be performed in the same way for each time step 
during the bubbling period, provided that the dataset is sufficiently large.

\section{Numerics\label{sec_numerics}}

\subsection{Numerical scheme}

The incompressible Navier-Stokes equations are solved for both phases 
(liquid and gas) using a sharp-interface arbitrary Lagrangian-Eulerian 
finite element method \citep{cairncross2000finite,Heil2006}. 
To that end, the geometry of the channel is meshed with triangular 
Taylor-Hood elements \citep{taylor1973numerical}. 
For simplicity, however, the channel is assumed to 
be axisymmetric, i.e.\ the deviation from a circular cross-section, as 
depicted in \cref{fig_setup}\,B or \cref{fig_postprocessing}\,C, 
is not taken into account. 
Implications of this assumption as well as of the assumption of
incompressible liquids will be addressed in the discussion section.
During the temporal integration, the mesh is always kept aligned with the 
moving liquid-gas interface, which requires the mesh
to move together with the phase boundary. This is achieved by treating the 
mesh as a pseudo-elastic body, which is deformed 
according to a displacement imposed 
via Lagrange multiplier fields at the moving interface 
\citep{cairncross2000finite,Heil2006}. The Lagrange 
multipliers ensure that the kinematic boundary condition is satisfied on both 
sides of the interface. When the mesh quality starts to suffer from the deformation, 
i.e.\ a single element either has doubled or halved its area with respect to the initial area, 
or the interior angles of the element drops below a threshold, the mesh is reconstructed and the pressure 
and velocity fields are subsequently interpolated to the new mesh.
The implementation is performed using the open-source finite element library 
\textsc{oomph-lib} \citep{Heil2006}. 
The general model implementation has been successfully validated with 
simulations for a variety of other physical phenomena, 
ranging from evaporation \citep{Li2020}, droplets bouncing in a stratified 
liquid due to buoyancy \citep{Li2019} and Marangoni flow and Leidenfrost 
droplets hovering on a bath \citep{Gauthier2019}.

While a benefit of the sharp-interface method is that the inclined walls at 
the outflow of the channel can be easily implemented, the topological changes that
both phases undergo during the pinch-off events require a more sophisticated 
approach as, e.g., in volume-of-fluid
methods.
Whenever the diameter of the gas domain falls below a 
critical threshold (here 2.5\,\% of  $w$ in \cref{fig_setup}\,\textbf{A}) at the imminent 
pinch-off point (local minimum in the gas domain diameter), 
the liquid-gas interface is dissected and both ends are 
reattached to the axis of symmetry, while ensuring volume conservation. 
The mesh is then rebuilt as described above.

The density in the simulation was set to
\SI{997}{\kilogram\per\meter^3} for the liquid and \SI{1.25}{\kilogram\per\meter^3} 
for the gas. Dynamic viscosities were set to \SI{1}{\milli\pascal\second} 
and \SI{0.0175}{\milli\pascal\second} for
the liquid and gas phase, respectively.
Considering the microsecond timescale of bubble formation,
we assume that the Tween~80 used in the experiments does not 
have sufficient time to significantly cover the interface during bubble pinch-off
and the surface interfacial 
tension was therefore set to \SI{72}{\milli\newton\per\meter}.
Indeed, no significant differences in bubbling frequency and size have been 
experimentally observed between bubble production with pure water and water-Tween solution.
In the experiment, the surfactant thus mainly increases the stability 
of the bubbles against coalescence through its presence between the colliding bubbles
in the expanding outlet channel, as previously shown for phospholipids \citep{segers2017universal}.

A Poiseuille-like Dirichlet velocity profile was imposed at 
the liquid inlet and the volumetric inflow can be set to any desired flow rate. 
The gas is pushed into the channel via a normal traction boundary 
condition to impose the chosen gas over-pressure. 
The outflow is an open, normal-stress-free, boundary.

\subsection{Validation of the numerical model}

The experimental channel geometry is described in 
\cref{fig_setup}\,A and corresponds to a hydraulic diameter of
\SI{18.4}{\micro\meter}. Consequently, for the numerical
simulation with a circular cross-section
a diameter of \SI{18.4}{\micro\meter} was chosen. 
Furthermore, with this choice, the channel cross-sectional area is 
conserved within 3\,\% accuracy.
Due to the differences in the precise channel geometry
and due to the incompressibility of the simulation, some 
quantitative variations between experiments and simulations can be 
expected when considering a large range of parameters.
However, as we will show in the following, simulations and experiments 
agree well for the parameter settings considered in this paper. 
We validate the numerical results using two specific cases.

\begin{figure*}
 \includegraphics{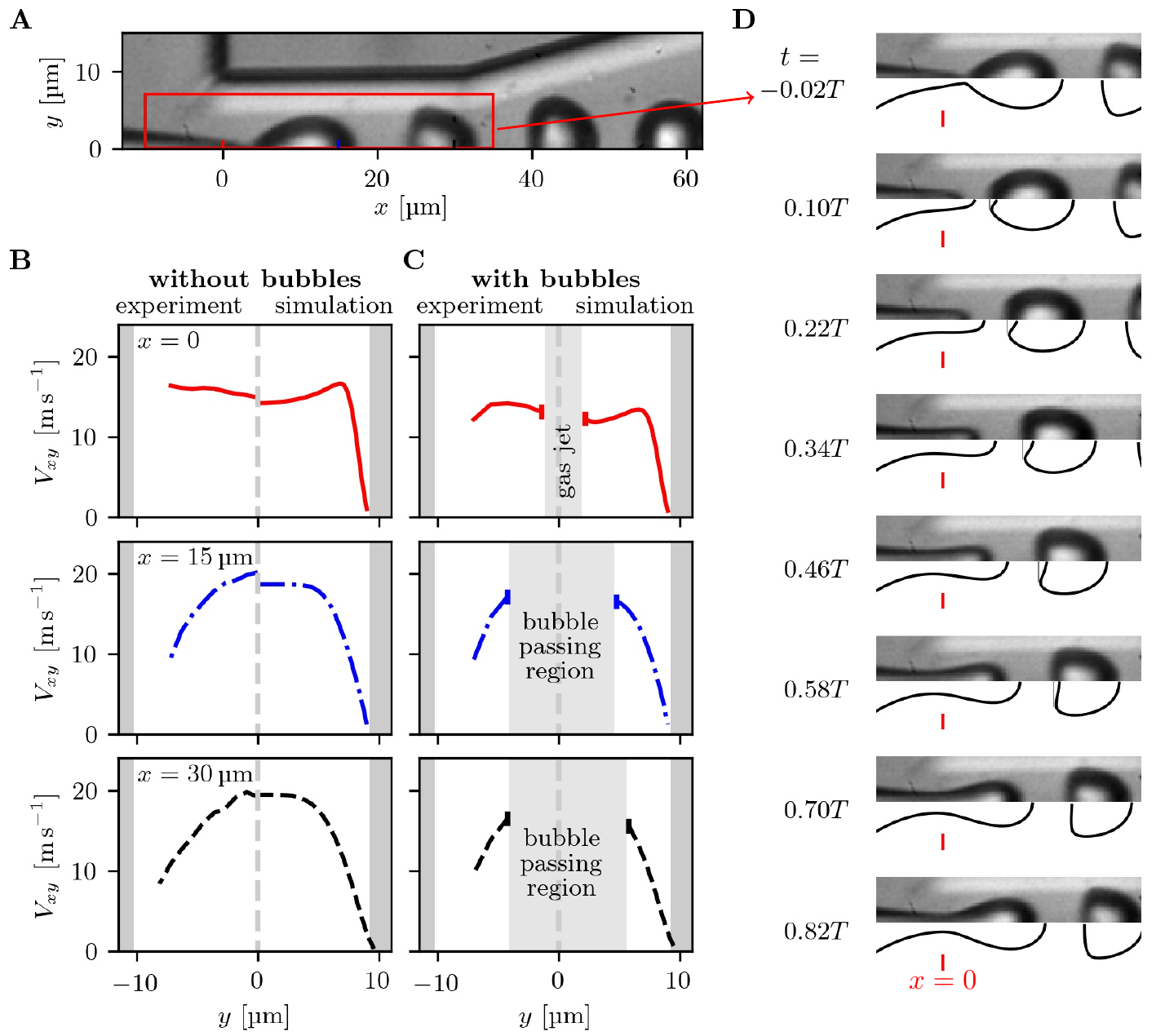}%
  \caption{Comparison between numerical model and experimental results.
  \textbf{A} -- Snapshot for reference. 
  \textbf{B} -- Average flow profiles for experiments and simulations
  without bubbles. 
  \textbf{C} -- Average flow profiles for experiments and simulations
  with bubbles. The center region where bubbles pass has been left 
  us on purpose in order not to induce any bias through averaging. 
  \textbf{D} Temporal evolution of the jet tip
  and detached bubble in the flow-focusing channel, experiments 
  (upper halves) compared to the numerically obtained gas-liquid interface
  (black line in the lower half plots). The thin grey lines that 
  are added at the rear of the numerically obtained bubbles
  indicates the shape that would result 
  from a projection of the bubble for comparison to the experimental shape.
\label{fig_numerics}}
\end{figure*}

Case (A): Liquid flow in the absence of bubbles at a flow rate of
\SI{0.2}{\milli \liter \per \minute} is considered as the reference.
Experimental and numerical curves are shown in \cref{fig_numerics}\,B.
This case allows to compare the experiments and the numerics 
to the analytical solutions for pipe flow.
It is generally assumed that the flow develops from a flat profile
in the inlet, to a parabolic profile corresponding to the well-known
Hagen-Poiseuille solution.
In \cref{fig_numerics}\,B, the profile for $x\approx 0$
is not perfectly flat, but 
presents a maximum close to the channel wall, a phenomenon reported
analytically e.\,g.\ by \citet{goldberg1988solutions} 
and numerically by \citet{dombrowski1993influence}.
The transition from a flat to a parabolic profile occurs over a typical distance
defined as the entry length. For Reynolds numbers similar to the ones
used here, $\mathrm{Re}\approx 400$,  
\citet{dombrowski1993influence} report an entry length of approximately 
twenty times the inlet diameter.
Indeed, in the numerical case, it is visible that the parabolic shape
is not fully reached, while the experimental results show a fully 
developed profile after a distance of merely $1.5$ times the channel diameter. 
The difference between experiments and simulations may be 
due to slightly larger radial velocity components and irregularities 
in the channel geometry.

Case (B): 
The liquid flow is set to \SI{158}{\micro \liter \per \minute} 
and the numerical gas pressure is set to \SI{1.4}{\bar}.
This leads to the production of bubbles with a radius  
$R_b = \SI{5.5}{\micro\meter}$  formed at a rate of 
$1.22\cdot 10^6$ bubbles per second and traveling at approximately \SI{20}{\meter \per \second}.
The gas flow rate can be calculated from the bubble size and bubbling
frequency and gives
$Q_g = \SI{40}{\micro \liter \per \minute}$.
Adding the liquid flow rate yields a total of 
$Q=Q_l + Q_g \approx \SI{200}{\micro \liter \per \minute}$.
Thus, the total flow rates for cases (A) and (B) are comparable.
This specific numerical setting provides a solution which 
is quantitatively comparable with the 
experiments when the chip is driven at a gas pressure of \SI{2.7}{\bar}.
The difference in the set backing pressure can be explained by longer 
tubing and hence an increased total pressure drop in the experiments.
Experimentally, $R_b$ ranges from $4.5$ to $\SI{6.0}{\micro\meter}$, where
the bubble size inside the channel is oscillating due to 
compressibility effects and due to small variations between different recordings.
\Cref{fig_numerics}\,C shows the comparison between 
velocity profiles in the liquid phase, where the markers denote
the edge of the bubbles.
\Cref{fig_numerics}\,D shows good agreement in bubble size, 
shape and position between experiments and simulations for case~(B).
The largest difference concerns the size of the gas jet upstream of the channel 
entrance, i.e.\ for $x<0$.
This difference may be explained by the fact that in the experimental case, the 
liquid streams through the two supply channels into the flow-focusing region,
whereas the numerical case is axisymmetric resulting in an equally distributed
inflow from all directions.
The reference time $t=0$ corresponds to the moment of pinch-off.

\section{Results}

\subsection{Average fields\label{sec_average}}

To understand the role of the flow and pressure components 
on the dynamics of the flow-focusing nozzle, it is convenient 
to separate their contributions in time-averaged fields and in 
oscillatory components. In this section we will first investigate the 
influence of bubbles on the average fields.

\subsubsection{Time-averaged velocity field}

\begin{figure*}
\includegraphics{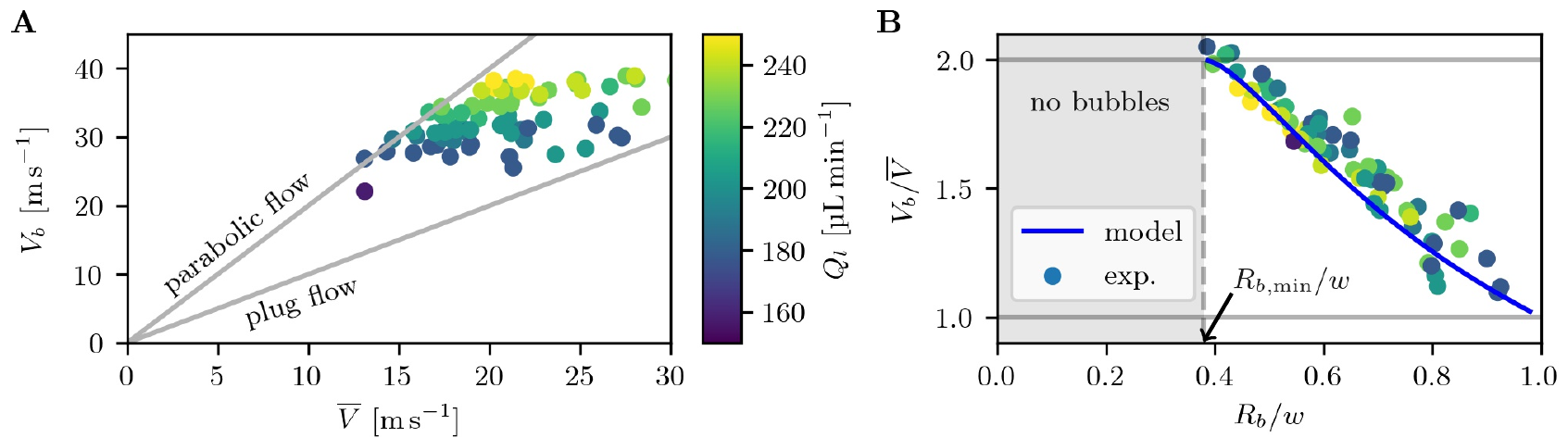}
\caption{\textbf{A} -- Experimental bubble velocity $V_b$ 
at the end of the flow-focusing channel
as a function of the corresponding mean flow velocity
$\overline{V} = (Q_g + Q_l) / A$ for a large number of experimental 
bubble sizes and liquid flow rates.
The color code shows the different liquid flow rates
used during bubble production.
\textbf{B} -- Flow rate ratio $V_b/\overline{V}$ as a function of
the bubble radius normalized by the channel radius and corresponding model from
\cref{eq_flowprofile,eq_boundaryconditions,eq_int,eq_constants,eq_correctionR}.
No bubbles smaller than $R_{b\mathrm{,min}}$ could be formed in the present
flow-focusing device and within the parameter space explored.
\label{fig_ProfileType}}
\end{figure*}

A detailed look at the flow profiles in 
\cref{fig_numerics} 
reveals that the larger the bubbles, 
the steeper the velocity gradient near the channel wall. 
A first simple approach for the developed flow is to consider plug flow in the 
center of the channel at the location of the bubbles and to
assume a solution of the Navier-Stokes equation for the 
region between the bubbles and the wall:
\begin{equation}\label{eq_flowprofile}
	V_x(r) = \left\{\begin{array}{ll}
	 V_b \hspace{10pt} \text{for}  \hspace{5pt} r \leq R^\ast, \\
	 - \frac{K r^2}{4} + A \ln\left(\frac{r}{w}\right) + B 
	 	 \hspace{10pt} \text{for}  \hspace{5pt}  R^\ast \leq r \leq w.
	\end{array} \right.
\end{equation}
Here, $w$ is the channel width, $r$ the radial coordinate and 
$R^\ast$ denotes the location of the transition between a parabolic and a flat profile.
All variables of length ($r$, $w$ and $R^\ast$) are in units of meter.
The unknown velocity of the bubble $V_b$ as well as the constants 
$K$, $A$ and $B$ can be found using the boundary conditions,
\begin{subequations}\label{eq_boundaryconditions}
\begin{align}
V_x(r=w) &=0 \hspace{5pt}, \\
V_x(r=R^\ast)&=V_b \hspace{5pt}, \\
\frac{\mathrm{d}}{\mathrm{d}r}V_x(r=R^\ast) &= 0 \hspace{5pt}, \label{eq_boundaryconditionsC}
\end{align}
\end{subequations}
and the integration 
\begin{equation}\label{eq_int}
Q_t = \int_{r=0}^w 2 \pi r V_x(r) \mathrm{d}r \hspace{5pt},
\end{equation}
where $Q_t =Q_l+Q_g$ is the known total (i.e.\ liquid plus gas) flow rate.
\Cref{eq_boundaryconditionsC} implies a free shear boundary condition
and requires a negligible influence of surfactants on the liquid-gas boundary.
Even though typical adsorption times for convective models are of the order
of seconds \citep{bkak2016interfacial,chang1995adsorption} and thus
much larger than the present microsecond scale bubble production,
typical times for a convective process as presented here are not commonly known.
For this reason, the free shear boundary condition can also be
explained by an expected light packing of surfactants 
at an early stage of bubble production in view of its final packing.
In fact, it is known that a contrast agent bubble usually shrinks by a factor of approximately 2.5 
\citep{segers2016stability} reducing the surface area to 15\,\% of the
initial size. 
Consequently the surface is 85\,\% free upon formation and
the molecules should be highly mobile,  
and thus we can assume that shear effects for such a lightly packed
bubble are negligible \citep{marmottant2005model}.
The result from \cref{eq_flowprofile,eq_boundaryconditions,eq_int} yields 
\begin{subequations}\label{eq_constants}
\begin{align}
K&=\frac{8Q_t}{\pi (w^2-R^{\ast 2})^2} \hspace{5pt}, \\
A&=\frac{K}{2} R^{\ast 2} \hspace{5pt}, \\
B&= \frac{K}{4} w^2 \hspace{5pt}, \\
V_b&=\frac{K}{4} \left(w^2 - R^{\ast 2} + 2 R^{\ast 2} \ln\left(\frac{w}{R^\ast}\right)\right) 
\hspace{5pt}.\label{eq_vb}
\end{align}
\end{subequations}
Even though \cref{eq_vb} is not defined for the limiting cases,
it tends towards a parabolic profile $V_b = 2\overline{V}$ for $R^\ast \rightarrow 0$
and a flat profile $V_b = \overline{V}$ for $R^\ast \rightarrow w$, where
$\overline{V}=(Q_t/S)$ is the average velocity with the cross-sectional area
$S=\pi w^2$.
Both limiting cases are represented by the gray solid lines in 
\cref{fig_ProfileType}\,A,
which shows the velocity of the bubbles versus the average flow velocity.
The flow rate ratio as a function of the normalized 
bubble size is plotted in \cref{fig_ProfileType}\,B.
It can be observed that, within the parameter space explored,
the experimental data collapses on 
a single curve irrespective of the liquid flow rate applied.
In addition, the blue line shows the theoretical considerations from 
\cref{eq_flowprofile,eq_vb}.
Here, the assumption of 
\begin{equation}\label{eq_correctionR}
R^\ast = \left(R_b - R_{b\mathrm{,min}}\right) 
	\left(\frac{1}{1-\frac{R_{b\mathrm{,min}}}{w}}\right) 
\end{equation}
has been taken, where $R_{b\mathrm{,min}}/w=0.38$
corresponds to the minimum bubble size observed experimentally, 
see \cref{fig_ProfileType}\,\textbf{B}. 
The expression of $R^\ast$ in \cref{eq_correctionR} means a
correction of the size of the plug flow area in the center of the
channel. It takes into account that the plug flow region is smaller than the
actual bubble size. 
This is in qualitative agreement with experimental and numerical
observations, even though the time-average in the bubble passing region,
see \cref{fig_numerics}\,C, has to interpreted with caution and is not
shown in the present paper.
Indeed, the time-dependent flow profile is much more complex and
\cref{eq_flowprofile,eq_boundaryconditions,eq_int,eq_constants,eq_correctionR} 
are only a time-averaged approximation. 
An interesting feature of the model is that for minimum size bubbles, a
parabolic average flow-profile is assumed, which is in agreement with
both the experimentally observed flow-profiles such as
presented in \cref{fig_numerics}\,B and the bubble velocities in \cref{fig_ProfileType}\,B.

\subsubsection{Time-averaged pressure field\label{sec_meanPressure}}

The local pressure field is directly related to the flow velocity 
field via the Navier-Stokes equations and arises either from 
conversion of kinetic energy into pressure (volumetric potential energy) 
or from viscous losses. The derivatives of the velocity in the
convection and diffusion terms of 
these equations, however, do not allow to accurately recover the 
pressure field from the velocity measurements, owing to measurement noise. 
The pressure is therefore extracted from the numerical model.
\begin{figure*}
\includegraphics{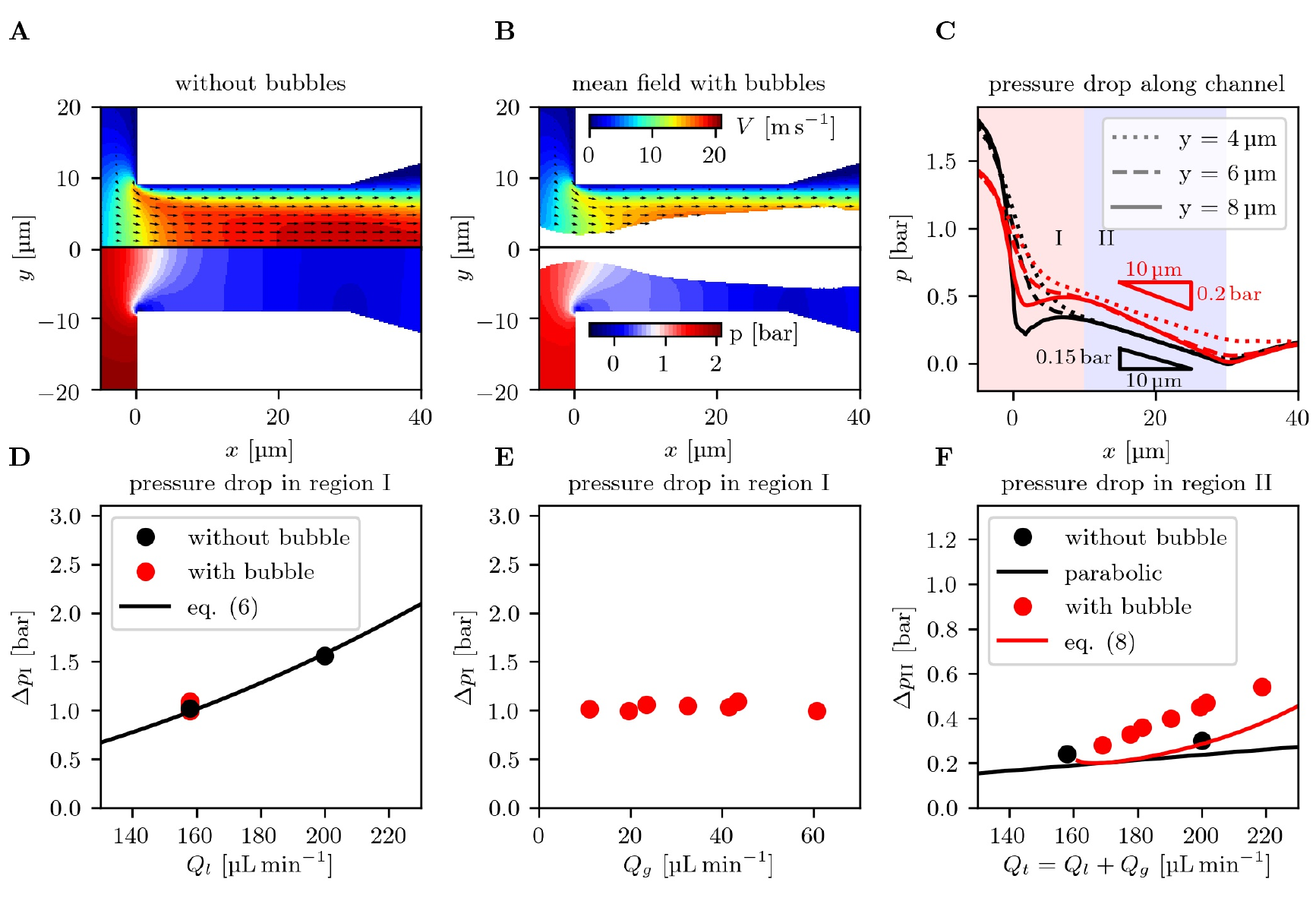}
\caption{
\textbf{A} and \textbf{B} -- Average velocity (upper half) and pressure fields (lower half)
for the case~(A) without bubbles and the case~(B) with bubbles, respectively.
Note that for part \textbf{B} the area where bubbles pass has intentionally 
been left blank in order not to induce any ambiguities in the definition
of the mean velocity value.
\textbf{C} -- Average pressure along
three different lines parallel to the $x$-axis without bubbles (black line) 
and with bubbles (red line), both 
at a total flow rate $Q_t = Q_l+Q_g=\SI{200}{\micro\liter\per\minute}$.
Note that for the radial position $y=\SI{4}{\micro\meter}$ for the case with bubbles 
the time-average pressure includes moments where the gas phase is present,
which leads to a slightly increased average pressure.
As a reference, the minimum pressure appearing at about the channel end at
$x\approx\SI{30}{\micro\meter}$ was set to zero.
\textbf{D} and \textbf{E} -- Pressure drop in the inlet region I as a function of liquid flow rate and
gas flow rate respectively.
\textbf{F} -- Pressure drop the region II as a function of total flow rate.
\label{fig_PressureEmpty}}
\end{figure*}
The corresponding  pressure and velocity fields for 
case~(A) without bubbles
and case~(B) with bubbles are shown in figs. \ref{fig_PressureEmpty}\,A and B,
respectively. The pressure profiles along the axial position $x$
are plotted in 
\cref{fig_PressureEmpty}\,C 
for three radial positions $y=4$, $6$ and $\SI{8}{\micro\meter}$.

In the inlet region of the flow-focusing channel, defined here
by $-\SI{10}{\micro\meter}<x<\SI{10}{\micro\meter}$ and indicated
as region I in \cref{fig_PressureEmpty}\,C
the liquid experiences a sharp pressure drop,
which arises from both, Bernoulli effects ($\Delta p_\mathrm{dyn}$)
and viscous entry losses ($\Delta p_\mathrm{loss}$),
\begin{equation}\label{eq_losses}
\Delta p_\mathrm{I} 
	= \Delta p_\mathrm{dyn} + \Delta p_\mathrm{loss}
	\approx \frac{1}{2} \rho \overline{V}^2 
	+ K_i\frac{1}{2} \rho \overline{V}^2 \dist,
\end{equation}
where $K_i$ the coefficient of resistance associated with the geometric shape 
of the inlet region, a number usually 
obtained empirically \citep{steinke2006single}.
The velocity in the large channels upstream of the flow-focusing channel
is much lower than in the narrow flow-focusing channel itself and, therefore,
it is neglected in the definition $\Delta p_\mathrm{dyn}$.
For case (A) without bubbles
and with a liquid flow rate of $\SI{200}{\micro\liter\per\minute}$,
$\Delta p_\mathrm{dyn} = \SI{0.76}{\bar}$,
for case (B) with bubbles 
and a liquid flow rate of $\SI{158}{\micro\liter\per\minute}$,
$\Delta p_\mathrm{dyn} = \SI{0.47}{\bar}$.
In both cases, this accounts for about half of the total pressure decrease 
(black points in \cref{fig_PressureEmpty}\,D)
in the inlet region~I, 
the other half is attributed to viscous losses in the inlet region 
with $K_i\approx 1$. 
Strikingly, the pressure drop in region I, $\Delta p_\mathrm{I}$, 
does not significantly depend on the bubble size,
see red points in \cref{fig_PressureEmpty}\,D and~E. 
This suggests that, in the inlet region I, 
the gas jet has very little influence on the total flow behavior. 
The liquid flow rate is therefore the only relevant parameter
to quantify the pressure drop in region~I.

One can see from \cref{fig_PressureEmpty}\,C that the
pressure loss in region II
is linear, which is consistent with the hydraulic pressure drop for a Poiseuille flow,
\begin{equation}\label{eq_pdroplinear}
\Delta p_\mathrm{linear} = R_H q_l \dist,
\end{equation}
where $R_H = (8 \mu L)/(\pi w^4)$ is the hydraulic resistance \citep{bruus2008theoretical}.
The theoretical pressure drop along the $\SI{20}{\micro\meter}$ channel length considered here
is plotted by the black solid line in \cref{fig_PressureEmpty}\,F for the 
case without bubbles. The theory slightly underestimates the 
numerically obtained pressure drop (black dots for simulations without bubbles).
An explanation can be found in the slightly larger pressure gradient
near the wall for the undeveloped flow in the simulation.
In the case with bubbles, numerically obtained red dots in
\cref{fig_PressureEmpty}\,F, the underestimation with respect to the theoretic 
estimation of a parabolic flow is even stronger.
To take into account the deformation of the flow field as discussed in the above model,
the pressure drop can be calculated to be
\begin{equation}\label{eq_pressuredrop}
\frac{\mathrm{d}p}{\mathrm{d}x} 
= \eta \frac{1}{r} \frac{\partial}{\partial r} \left( r \frac{\partial V_x}{\partial r} \right)
= \frac{K}{\mu} = -\frac{8 \mu Q_t}{\pi (w^2 -R^{\ast2})^2} \hspace{5pt}.
\end{equation}
The result is shown in by the red line in \cref{fig_PressureEmpty}\,F.
The remaining mismatch can mostly be attributed to 
a steeper velocity gradient close to the walls in the simulation
for the undeveloped flow.

\FloatBarrier

\subsection{Oscillatory fields\label{sec_oscillations}}

Oscillatory fields are of particular interest here since
they potentially result in cross-talk between parallelly connected nozzles. 
It is therefore essential to quantify the oscillations and  
understand their origin.

\subsubsection{Oscillations in the gas phase\label{sec_gas}}

\begin{figure*}
\includegraphics{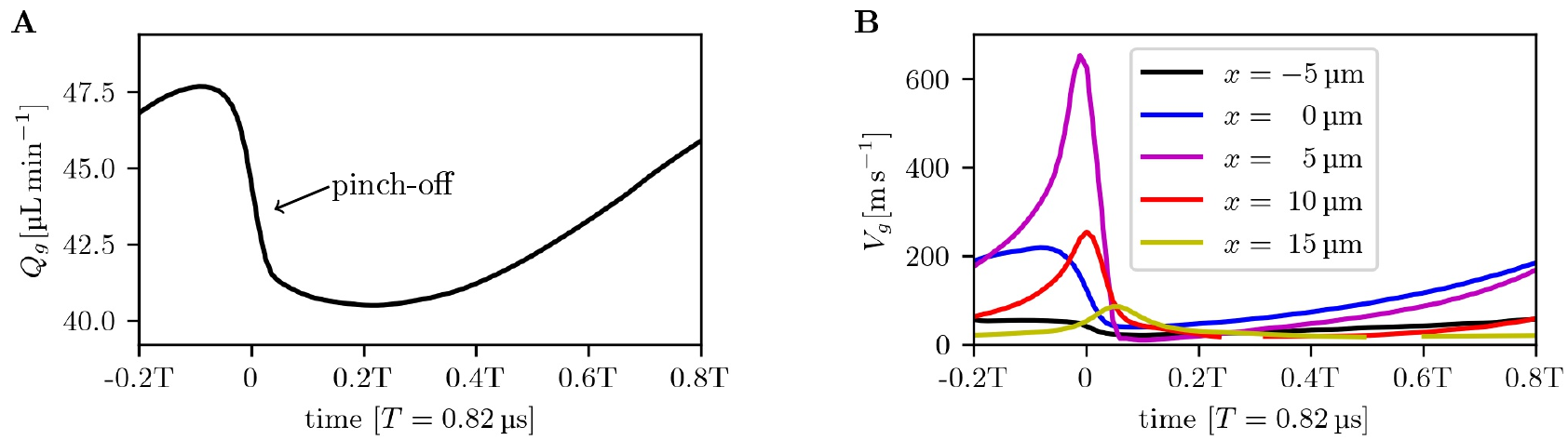}
\caption{\label{fig_gas}Numerical oscillatory velocity in the gas phase
for $y=0$ and different positions $x$ in the flow-focusing channel. 
\textbf{A} -- gas flow rate as a function of time.
\textbf{B} -- gas velocity for different positions inside the gas thread 
(position of pinch-off is $x\approx\SI{5}{\micro\meter}$).
Limitations of the incompressible simulations are discussed in the text.}
\end{figure*}

Oscillations in the gas phase will have a direct impact on the
liquid velocity inside the flow-focusing channel.
The employed method of backlighting, where bubbles are imaged as black regions,
makes it however impossible to use
tracer particles to image the velocity of the gas jet.
We therefore have to rely on our numerical model. 
\Cref{fig_gas}\,A shows the gas flow rate which is directly related
to the gas velocity inside the gas supply channel.
In this pressure-controlled system, the gas phase experiences velocity 
oscillations \citep{wilkinson1994theoretical,gordillo2007bubbling}
and consequently flow rate oscillations, here about 20\% of the average value.
To understand the origin of these gas flow oscillations,
hydraulic losses and Laplace pressure (see also section~\ref{sec_PressureOscillations}) 
need to be taken into account.
After the moment of pinch-off, there is temporarily no jet neck,
but a forward moving jet tip with a newly forming bubble. 
The Laplace pressure inside the bubble is decreasing in the beginning
leading to an increased gas flow rate until 
hydraulic pressure losses become important and finally the new
jet tip leads to a sudden increase of Laplace pressure,
both decreasing the gas flow rate.

Around the position of pinch-off the 
thinning gas jet with a periodically variable diameter increases both
the average gas velocity and the overlying oscillations, see \cref{fig_gas}\,B, leading to
velocities of several hundred meters per second.
While it is clear that such high velocities imply that our incompressible 
simulations do not completely represent the experimental case, 
supersonic flow velocities have been observed experimentally before
\citep{gekle2010supersonic}, and
we believe that our simulation provides sufficient qualitative insight into
the predominant effects. Due to the compressibility
of the gas before the jet neck, we can expect to observe lower
flow velocities in the experimental case.
Further aspects of the compressibility on the bubble dynamics
will be addressed in the discussion section.

\subsubsection{Oscillatory velocity field}

\begin{figure*}
\includegraphics{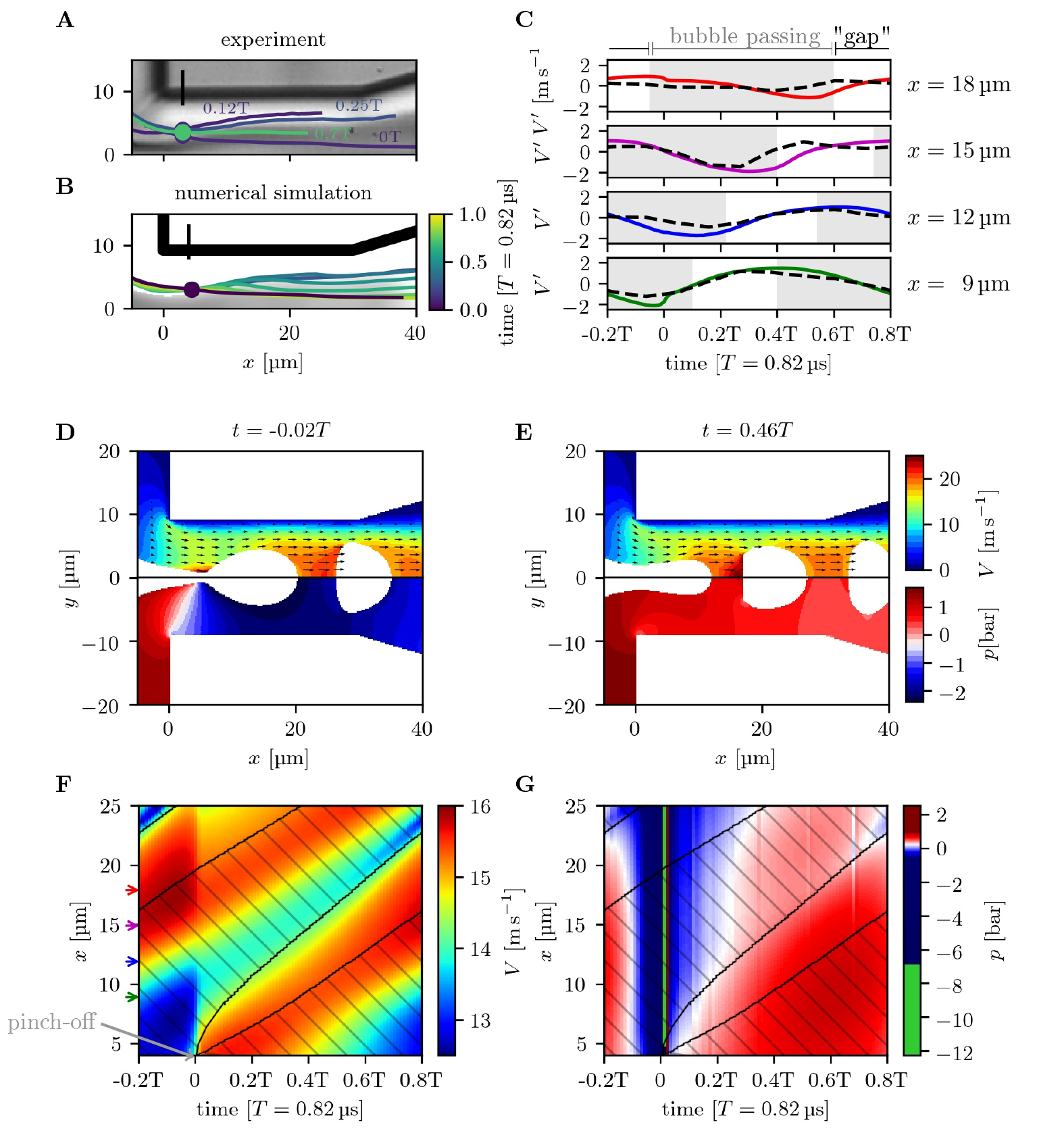}
\caption{\label{fig_voscillations}
Velocity and pressure oscillations in the flow-focusing channel.
\textbf{A} and \textbf{B} -- Typical experimental and numerical particle 
trajectories highlight the time-dependent 
behavior at a given location.
The time-stamps correspond to the moment when the respective 
particles were passing the position indicated by the dots.
\textbf{C} -- Velocity oscillations $V' = V-\overline{V}$
for positions $y=\SI{5}{\micro\meter}$ and $x$ given in the figure.
The colored lines present the numerical results and the dashed black line the
corresponding experimental data.
\textbf{D} and \textbf{E} -- Velocity and pressure map for two instants
during the bubbling period $T$.
\textbf{F} and \textbf{G} -- Velocity $V$ and pressure $p$, respectively,
for a fixed radial position $y=\SI{6}{\micro\meter}$,
different positions $x$ and different moments in time. 
The hatched areas corresponds to the presence of a bubble in the 
respective channel sections~$x$. The colored
arrows in \textbf{F} indicate the positions in \textbf{C}.}
\end{figure*}

Gas flow oscillations and the periodic generation 
of bubbles may lead to oscillations of the liquid 
velocity field as well. A simple way of exemplifying the unsteady
nature of the liquid flow are the intersecting particle trajectories in
\cref{fig_voscillations}\,A and B, 
for experiments and numerical simulations, respectively. 

The source of the phenomenon of intersecting particle trajectories
becomes apparent when quantifying the flow field variations 
at the timescale of bubble generation. Velocity 
fluctuations in the continuous phase are plotted in 
\cref{fig_voscillations}\,C at four downstream locations in the channel
and the radial position $y=\SI{6}{\micro\meter}$, 
both for the experiments and for the simulations. 
For these positions, the velocity oscillates by $1$ to  
$\SI{2}{\meter\per\second}$ for mean flow velocities of 
the order of $\SI{15}{\meter\per\second}$. 
Furthermore, \cref{fig_voscillations}\,C 
also shows that the velocity oscillations propagate along 
the channel axis, pushed along with the microbubble.  
A more extended look at the simulations in \cref{fig_voscillations}\,F
reveals that in addition to the propagation along the channel axis,
the oscillations are influenced by the bubble pinch-off at $t= 0$ as well.
\Cref{fig_voscillations}\,F shows the velocity (color) in the flow-focusing channel
for a fixed radial position $y=\SI{6}{\micro\meter}$ as a function
of both the axial position $x$ and time $t$.
The hatched areas indicate the presence of a bubble in the center of the channel.
\Cref{fig_voscillations}\,F highlights that the region of increased velocity
coincides with the front of the bubble.
Furthermore, note that strong variations are present at the moment
of bubble pinch-off, i.e.\ at $t=0$.

\subsubsection{Oscillatory pressure field\label{sec_PressureOscillations}}

The neck shape is the origin of hydraulic losses 
and Laplace pressure. 
The pressure in the liquid can thus be seen as a consequence of the neck shape.
The pressure inside the bubble is furthermore related to the pressure in the liquid via
the Laplace pressure of the bubble.
Orders of magnitude at the final stage before pinch-off (e.g. assuming a
\SI{0.1}{\micro\meter} neck radius) are a \SI{7}{bar} pressure difference
between liquid and gas at the minimum neck radius 
due to Laplace pressure and a \SI{0.35}{bar\per\micro\meter} hydraulic pressure loss along 
the gas thread.
In that sense the pressure impulse originates from the final, very short moment
with a rapidly increasing Laplace pressure of the jet neck.
After the moment of pinch-off, there is temporarily no jet neck,
but a forward moving jet tip with a newly forming bubble. 
Thus, it is the Laplace pressure of the bubble that is dominating,
however it decreases with increasing bubble size.
This also explains the oscillations of the gas velocity:
The Laplace pressure inside the bubble is decreasing in the beginning
leading to an increased gas flow rate until 
hydraulic pressure losses become important where finally the new
jet tip leads to a sudden decrease of the gas flow rate.

Oscillating gas and liquid velocities are bound to 
give rise to pressure oscillations. 
Figs.~\ref{fig_voscillations}\,D and E 
show the pressure field at two specific moments of the 
bubble generation process, before and after 
bubble pinch-off (refer to \cref{fig_numerics} for the exact time stamps).
The pressure varies by approximately $\SI{3}{\bar}$ between these two moments. 
\Cref{fig_voscillations}\,G displays the pressure 
for a fixed radial position $y=\SI{6}{\micro\meter}$,
different positions $x$ and different moments in time. 
One can in particular notice a 
strong pressure decrease around the moment
of bubble pinch-off. 
This pressure drop is a consequence of the neck shape and
Laplace pressure of the neck, which is increasing with decreasing jet radius
(i.e. \SI{0.1}{\micro\meter} neck radius leading to \SI{7}{bar} Laplace pressure).
Through continuity and a relatively weak Laplace pressure of the bubble before 
pinch-off, the gas pressure inside the bubble also experiences a large pressure
drop, which can be further explained by the hydraulic losses due to the
very large gas velocities in the thinning jet.
Between the bubble pinch-offs (here the time span $t=0.2T$ to $0.8T$),
pressure values between $0$ and $\SI{1}{\bar}$ are observed in the channel.
As for the average pressure, a general decrease along the $x$-axis is visible
due to viscous losses already highlighted for the average field in 
\cref{fig_PressureEmpty}.
Furthermore, it can be observed that the region around the newly forming
bubble corresponds to a region of higher pressure 
(bottom right area in \cref{fig_voscillations}\,G).
The use of an incrompressible solver can be expected to have an influence on
the exact shape and amplitude of the pressure pulse around $t=0$.
If one accounts for compressibility, especially locally, 
the impulse will probably be less strong, and wider in time, 
thus effectively spreading the energy over a longer timescale.
We believe, however, that the qualitative result of a strong pressure
peak is representative for the real phenomenon.
The phenomenon of acoustic emission at bubble pinch-off has been discussed in literature
theoretically \citep{longuet1990analytic} and numerically \citep{oguz1991numerical} 
in order to understand the sound of rain, and later also for the 
pinch-off of bubble trains \citep{manasseh2001passive,liu2018numerical}.
Among the different possible sources of sound proposed by \citet{oguz1991numerical},
the most likely candidates for the large impulses at the moment of pinch-off
observed in our simulations are the difference in Laplace pressure and the radial liquid inrush.

\section{Discussion\label{sec_discussion}}

In the result section we present results on the average and oscillatory 
components of the velocities and pressures inside the flow-focusing device.
While satisfactory agreement is obtained for the liquid velocities
between the experimental and numerical method, it is important to
keep in mind their respective limits.
For the experimental results, we estimate the error on 
the absolute liquid velocities to be in the range $\pm 2\dots\SI{3}{m/s}$.
This results mainly (1) from small perturbations during the operating time
of more than an hour, (2) from the particle tracking algorithm which occasionally 
outputs wrongly connected particle positions and therefore incorrect velocities
which increase the number of outliers, and (3) from the need to extract velocities
via an envelope due to detection of particles in the entire channel.
Nonetheless, relative velocity oscillations can be detected with higher accuracy.

Numerically, the main limitation arises through the use of an incompressible 
solver. As already discussed in section~\ref{sec_gas},
this leads in particular to an overestimation of the gas flow velocities
before the position of pinch-off. 
The simulated incompressibility of the gas also leads to a constant bubble volume 
once the bubble is pinched off, while in the experimental case the bubble
radius is oscillating by about 10\,\% while traveling down the flow-focusing channel.
This type of dynamics is known from the Rayleigh-Plesset type description for a free bubble
\citep{lauterborn1976numerical}. In the present study, however, it is further
influenced by the specific liquid-gas interaction confined in the channel.
Another important limitation of the incompressible solver for both the
liquid and gas phase is a infinite speed of sound and
consequently the lack to correctly represent acoustic effects.
In the real scenario, low compressibility effects give rise to the wave equation, 
and thus to the propagation of an acoustic impulse.
Due to the small scale of the chip, the time to propagate through the 
\SI{20}{\micro\meter} long channel would still be very short  $\sim\SI{15}{\nano\second}$.
The time to travel between several parallelized channels would
be of the order of a few hundred \si{\nano\second}.

Lastly, it should be noted that both the completely 
circular numerical cross section and the flattened experimental 
cross section differ from other typical channel types, in particular those with rectangular cross sections. 
Calculating the pressure drop in rectangular channels is more complex 
(see \citet{bruus2008theoretical} for more details) and taking into account bubbles 
is expected to be less straightforward than for a circular cross section. 
Especially if the bubble nearly fills the complete channel, the effect of 
flow in the sharp corners becomes important \citep{van2009flows}. 
In general, however, the influence of the corner flow gets weaker 
towards the center of the channel, where our small bubbles are passing \citep{moharana2013generalized,bruus2008theoretical}. 
We thus believe, that our results will also apply to other channel geometries, 
provided that the channel aspect ratio remains of the order of one and that bubbles 
remain small compared to the channel size.
\bigskip

Our results highlight the oscillatory nature of the velocity and pressure 
inside a flow-focusing device, where we observe oscillations in the gas flow 
rate by numerical simulation, in agreement with earlier work by 
\citet{wilkinson1994theoretical,gordillo2007bubbling}. Even though compressibility 
effects may lead to lower maximum velocities, values of the order 
of the speed of sound have been observed as shown before experimentally by 
\citep{gekle2010supersonic}. The collapsing jet neck leads to a pressure impulse as confirmed 
by numerical simulations. Once again, the exact pressure amplitude may differ from 
the compressible case, but our findings are in agreement with the observation of 
acoustic emission at the moment of bubble pinch-off found in other systems 
\citep{oguz1991numerical,longuet1990analytic,manasseh2001passive,liu2018numerical}. The oscillating gas 
phase leads to oscillations of the liquid velocity. To the best of our knowledge 
these oscillations have not been studied before.

Our study gives detailed insight into the physical processes linked to the bubble 
production inside a flow-focusing chip. It highlights how both, 
the average and the oscillatory field are influenced by the presence of the bubbles. 
These results can be used in further studies to understand potential communication 
between parallelized channels that could lead to a decreased monodispersity 
\citep{jiang2010mass,kendall2012scaled}. The insights on the dynamic flow fields 
investigated in this paper will therefore be instrumental to a successful design 
as to drive parallelized high-production rate nozzles for microfluidic industrial applications.

\section{Conclusions\label{sec_conclusion}}

The present study shows the time-averaged and time-resolved 
velocity field and pressure field in a flow-focusing device. 
Particle tracking has been used to evaluate the velocity field 
experimentally, and numerical simulations were used to provide 
insight in both the velocity and pressure fields. We can observe 
three types of oscillations. (1)~As we have a pressure-driven 
gas feeding, the gas velocity is oscillating as a result of the 
changing Laplace pressure at the tip of the jet upon bubble pinch-off. 
This leads to velocity maxima of the order of several hundred meters per second
in the jet neck region.
(2)~Oscillations 
can also be observed for the liquid velocity. They are closely 
linked to the presence of the bubbles, propagate with the 
velocity of the bubble and amount to up to 25\% of the total velocity
in the case studied here in detail. 
(3)~Pressure oscillations in the liquid phase are dominated by 
a strong negative pressure impulse linked to the periodic bubble 
pinch-off. Understanding these different types of oscillations 
will be a valuable asset to understand the role of cross-talk 
for parallelized flow-focusing nozzles. 
Experiments and/or 
simulations of such coupled systems will, however, be necessary to fully 
understand the importance of the respective oscillatory components
in a coupled device.
In addition to the results on the oscillatory flow, the present 
study provides insight into the effect of bubbles on the time-averaged 
liquid flow and pressure field inside a flow-focusing chip. The 
velocity profile can be approximated by a simple analytic 
expression while the presence of bubbles leads to a significant 
viscous pressure drop in the flow-focusing channel.

\section*{Acknowledgements}

This project is partly financed by 
Holland High Tech with a public-private partnership allowance 
in the top sector High Tech Systems and Materials (HTSM) and by
Bracco Suisse S.A. 
G.L. acknowledges funding from the 4TU Precision Medicine program 
supported by High Tech for a Sustainable Future,
a framework commissioned by the four Universities 
of Technology of the Netherlands.
T.S. acknowledges funding from the 
Max Planck - University of Twente Center for Complex Fluid Dynamics.

%

\end{document}